# The humidity sensitivity and mechanism of strontium-hexaferrite ceramics

Hui-Feng Zhang[a], Jian Li[b], Gang-Qin Shao[c*], Zhen-Sheng Gao[d], Jin-Hua Liu[e], H.-N. Girish[f]

State Key Laboratory of Advanced Technology for Materials Synthesis and Processing, Wuhan University of Technology, Wuhan 430070, China

[a]x2008_zhf@163.com, [b]lijian3011568@163.com, [c]gqshao@whut.edu.cn, [d]gzx5674829@126.com, [e]287478959@qq.com, [f]girishhn25@gmail.com

**Abstract:** The humidity sensitivity of *n*-type strontium-hexaferrite semiconductors, X-type ($Sr_2Co_2Fe_{28}O_{46}$) and Z-type ($Sr_3Co_2Fe_{24}O_{41}$), are investigated at room temperature. The contact and reactions between water and material surface are demonstrated considering the formation of oxide ion vacancy, singly-ionized / fully-ionized oxide ion vacancy, and the chemisorbed / physisorbed / condensed water. An impedance spectra model is proposed considering the material-electrode interface, interior material, grain boundary, absorbed water and Warburg response. The mechanism of electronic coupled with protonic conduction and humidity sensing mechanism are explained.

## 1. Introduction

Some hexaferrites, *e.g.* the Z-type $Sr_3Co_2Fe_{24}O_{41}$ ($Sr_3Co_2Z$), show magnetically induced dielectric constant change (MD) and magnetically induced ferroelectricity (ME) effects at a low magnetic field and room temperature (RT)[1, 2]. They attract much attention due to potential applications in future functional devices. Otherwise, the absorbed water on $Sr_3Co_2Z$ has been observed by our group to behave a negative differential resistance (NDR) effect, which might be misled as an intrinsic one[3]. This means the research for humidity sensitivity of hexaferrites may give us more enlightenment to MD / ME effects, even though our original intention does not focus on it. Some ceramics are used as impedance- (resistive-), capacitive-, piezoresistive- and magnetoelastic-types of humidity sensors. The impedance type consists of ionic and electronic types[4, 5]. In electronic semiconducting sensing materials, water chemisorption decreases or increases the electronic conductivity ($\sigma$) depending on the *p*- or *n*-type nature of semiconductors[6-8].

The $Sr_3Co_2Z$ hexaferrite is formed according the $RSTSR^*S^*T^*S^*$ sequence (the asterisk indicates blocks rotated 180° around *c*-axis), while X-type $Sr_2Co_2Fe_{28}O_{46}$ ($Sr_2Co_2X$) according $(RSR^*S^*_2)_3$[2, 9]. The conduction in hexaferrites may be explained by Verwey's hopping (interchange of electrons) mechanism[10, 11]. The electrons hopping is between homonymous ions in more than one valence state ($Fe^{3+} \leftrightarrow Fe^{2+}$), distributed randomly over different lattice sites. Hexaferrites structurally form hexagonal closed packed stacking of oxygen with cations at octahedral (6), trigonal-bipyramidal (5) and tetrahedral (4) sites (the numbers indicate the coordination number, CN). $Fe^{2+}$ ions formed during high-temperature sintering would occupy preferentially only octahedral sites, while $Fe^{3+}$ ions could occupy octahedral or tetrahedral sites[12]. This is related to the oxidation states and changeable ionic radii with different CN values: $r_{Fe^{3+}}(4) < r_{Fe^{3+}}(6)$, $r_{Fe^{3+}}(6) \approx r_{Fe^{2+}}(4)$, $r_{Fe^{2+}}(4) < r_{Fe^{2+}}(6)$ [13]. Thus the hopping between tetrahedral-tetrahedral sites does not exist[12].

In this paper, the humidity sensitivity of *n*-type strontium-hexaferrite semiconductors, such as impedance or resistance, hysteresis, recovery and response times with respect to relative humidity (RH) and frequency, are investigated at room temperature. The mechanism of electronic coupled with protonic conduction and humidity sensing mechanism are explained.

## 2. Experimental

The method for synthesizing powders of $Sr_2Co_2X$ (sample $X_M$) and $Sr_3Co_2Z$ (sample $Z_M$) was reported elsewhere[3, 14-18]. Ceramics were prepared by pressing powders into tablets ($\Phi 12 \times 3$ mm) and then sintering tablets at 1150 ℃ for 4 h (sample $Z_M$) and 1250 ℃ for 8 h (sample $X_M$) in $O_2$, respectively. The microstructure was examined by a field-emission scanning electron microscopy (FESEM, HITACHI S-4800, Japan). The tested density of sample $X_M$ (4.71 g/cm$^3$) is larger than that of $Z_M$ (4.29 g/cm$^3$) based on Archimedes principle. And the tested resistance of sample $X_M$ ($1.01 \times 10^{10}$ Ω·cm) is higher than that of $Z_M$ ($2.64 \times 10^9$ Ω·cm) by a bridge megger. The humidity characteristics were carried out by a CHS humidity intelligent analysis system (Elite, China) at RT. Before humidity testing, the sintered-tablets were polished to the size of $\Phi 10 \times 1$ mm, pasted with Ag slurry on both sides and fired at 830°C. The measured voltage was 1 V and frequency range was 10 Hz to 100 kHz. The controlled humidity environment was achieved by supersaturated aqueous solutions of $LiCl$, $MgCl_2$, $Mg(NO_3)_2$, $NaCl$, $KCl$ and $KNO_3$ at RT, which yielded 11%, 33%, 54%, 75%, 85% and 95% RH, respectively.

## 3. Results and discussion

### 3.1. Humidity sensing characteristics

Fig. 1 shows the dependence of impedance (Z) on relative humidity (RH) for sample $X_M$ and sample $Z_M$ at 10 Hz - 100 kHz. The insets are the corresponding SEM imagines. Here RH = $P_w / P_s \times 100\%$, $P_w$ and $P_s$ are the water vapor pressures at the actual value and saturation, respectively. It can be concluded that the impedance at various RH (the slope is defined as sensitivity) is strongly affected by the measuring frequencies, especially in relatively low range. While the RH varies from 11% to 95% at 10 Hz - 100 kHz, the impedance decreases by about three orders magnitude (~$10^8$ Ω to $10^5$ Ω) for sample $X_M$ and two orders magnitude (~$10^7$ Ω to $10^5$ Ω) for $Z_M$, showing a sensitive humidity response and good monotonicity in the entire humidity range pertaining to an *n*-type behavior.

For hexaferrite samples, the conduction mechanism originates from chemical and physical adsorption of water on their surface as well as the capillary condensation of water inside pores. Fig. 2 shows contact and reaction between water and semiconductor surface at different humidities.

(1) At low humidity (stage a to g in Fig. 2) where electronic conduction is dominant:

Excess iron, or oxide ion vacancy ($V_{\ddot{O}}$, Fig. 2a & 2b) during sintering at high temperatures, increases the proportion of $Fe^{2+}$ (Fig. 2c) and *n*-type conduction (which is very popular in ferrites). Referring to the work of Morimoto et al.[19], McCafferty et al.[20], Seiyama et al.[7] and Shimizu et al.[21], the following defect equations are suggested using Kröger-Vink notation:

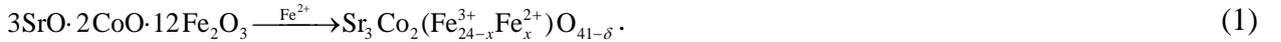

$$3SrO \cdot 2CoO \cdot 12Fe_2O_3 \xrightarrow{Fe^{2+}} Sr_3Co_2(Fe^{3+}_{24-x}Fe^{2+}_x)O_{41-\delta}. \tag{1}$$

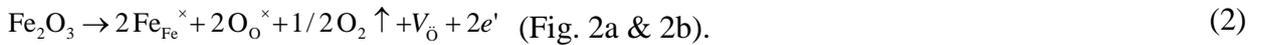

$$Fe_2O_3 \rightarrow 2Fe_{Fe}^{\times} + 2O_O^{\times} + 1/2O_2 \uparrow + V_{\ddot{O}} + 2e' \quad \text{(Fig. 2a \& 2b)}. \tag{2}$$

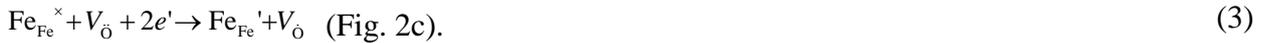

$$Fe_{Fe}^{\times} + V_{\ddot{O}} + 2e' \rightarrow Fe_{Fe}' + V_{\dot{O}} \quad \text{(Fig. 2c)}. \tag{3}$$

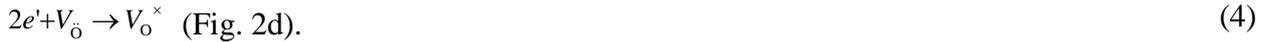

$$2e' + V_{\ddot{O}} \rightarrow V_O^{\times} \quad \text{(Fig. 2d)}. \tag{4}$$

where $Fe^{3+} = Fe_{Fe}^{\times}$, $O^{2-} = O_O^{\times}$ and $Fe^{2+} = Fe_{Fe}'$, $V_{\ddot{O}}$ stands for oxide ion vacancy, $V_{\dot{O}}$ stands for singly-ionized oxide ion vacancy (*i.e.* oxygen vacancy trapping one electron), and $V_O^{\times}$ stands for fully-ionized oxide ion vacancy (*i.e.* oxygen vacancy trapping two electrons)[7, 21].

Thus two sensing mechanisms can be deduced.

*Mechanism* I —— The reaction of a water molecule with a neutral oxygen site and a singly-ionized oxide ion vacancy provides one-electron donation.

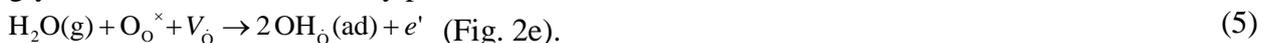

$$H_2O(g) + O_O^{\times} + V_{\dot{O}} \rightarrow 2OH_{\dot{O}}(ad) + e' \quad \text{(Fig. 2e)}. \tag{5}$$

Applying the law of mass action to Eq. 5 yields

$$K_1 = [OH_{\dot{O}}]^2[e']/([O_O^{\times}][V_{\dot{O}}]P_w). \tag{6}$$

Assuming that concentrations of both $[O_O^\times]$ and $[V_{\ddot{O}}]$ are almost constant and the electroneutrality principle ($[OH_{\dot{O}}] = 2[e']$) can be applied, the dependence of electronic conductivity ($\sigma$) on water vapor pressure ($P_w$) is obtained.

$$\sigma \propto P_w^{1/3}. \tag{7}$$

So the log-log plot of $\sigma$ vs. $P_w$ would give a straight line with a slope of 1/3.

*Mechanism* II —— The reaction of a water molecule with a fully-ionized oxide ion vacancy and a neutral oxygen site provides two-electron donation.

$$H_2O(g) + O_O^\times + V_O^{\times} \rightarrow 2OH_{\dot{O}}(ad) + 2e' \quad \text{(Fig. 2f)}. \tag{8}$$

$$K_2 = [OH_{\dot{O}}]^2[e']^2 /([O_O^\times][V_O^{\times}]P_w). \tag{9}$$

$$\sigma \propto P_w^{1/4}. \tag{10}$$

So the log-log plot of $\sigma$ vs. $P_w$ would give a straight line with a slope of 1/4 in this case.

At 100 Hz, the impedance changes little with RH for sample $X_M$ but inversely for $Z_M$. At higher frequency, the sensors show less sensitivity, especially for sample $X_M$. The impedance decreases effectively as the frequency increases. And the impedance difference at low humidity between two working frequencies is larger than that at high humidity. This effect is attributed to that the absorbed water at high frequency cannot be polarized, and the dielectric phenomena disappear[22]. The impedance is now controlled by the geometric capacitance of the sensor[23]. Since the observed impedance values were proportional to one-third power of the water vapor pressure for sample $X_M$ and one-fourth for $Z_M$, it is concluded that *Mechanism* I is for sample $X_M$ and *Mechanism* II for $Z_M$, respectively. The electronic conduction is dominant at low humidity even though $OH^-$ groups might dissociate to provide mobile $H^+$ through tunneling effect[6, 24].

(2) At median humidity (stage g to i in Fig. 2) where electronic coupled with protonic conduction (Grotthuss mechanism) is dominant:

With RH increasing, subsequent water is physisorbed orderly on the chemisorbed ($OH^-$) layer through two hydrogen bonds. Namely, it forms icelike clusters of hydrogen-bonded water molecules without hydrogen bonds between (not enough water). Thus no $H^+$ could be conducted in this stage (Fig. 2g)[6, 24]. Then a less-ordered extra layer on top of the 1st-physisorbed layer forms (Fig. 2h). There exists a high local charge density and a strong electrostatic field in the tips and surface defects, which can promote dissociation of the highly polar water physisorbed. $2H_2O \rightarrow H_3O^+ + OH^-$. Water molecules become gradually identical to bulk liquid water (Fig. 2i). $H_3O^+$ becomes the dominant charge carrier. Further, it is hydrated and release mobile $H^+$, $H_3O^+ \rightarrow H_2O + H^+$[24]. Each water molecule is singly bonded with one $OH^-$ group and the dominant carrier is $H^+$ (Grotthuss chain reaction[20, 25, 26]). Therefore, the conductivity of *n*-type strontium-hexaferrite semiconductors at RT is actually due to both intrinsic electrons conduction and Grotthuss proton-transfer. The resistance transduction principle can be expressed as[27-30]:

$$lg(R_{RH}/R_0) = (lga - lg(RH)^n)/(1 + b/(RH)^n). \tag{11}$$

where $R_0$ and $R_{RH}$ are the resistance at 0% and RH%, respectively. The $a$ and $b$ are the theoretical constants depending on the composition and pore structure of the ceramic. $(RH)^n$ is the concentration of quasi-liquid water[29, 30] representing the contribution of proton conductivity. And $n$ is the correction index. When $(RH)^n \rightarrow 0$, the Eq. 11 is corresponding to the Eq. 7 or Eq. 10.

(3) At high humidity (stage i to j in Fig. 2) where protonic coupled with electronic conduction is dominant:

The water is abundant now. At a given pressure, the multi-layers of physisorbed water molecules tend to condense into capillary pores with a radius below the Kelvin radius[25, 31, 32].

$$ln(RH) = ln(P_w/P_s) = -2\gamma V \cos\theta/(rRT) = -2\gamma M \cos\theta/(\rho rRT). \tag{12}$$

where $\gamma / V / M / \rho$ is the surface tension, molecular volume, molecular weight and density of water, respectively. $R$ is the gas constant, $T$ is the absolute temperature, $\theta$ is the contact angle between water and wall of capillary pore, $r$ is the Kelvin radius of capillary pore.

Microstructure may affect on the electrical resistivity and humidity sensitivity. Larger grains with less grain boundary and denser specimens imply better electrical conduction. And porosity benefits protonic conduction. Sample $X_M$ is less conductive than that of $Z_M$ ($1.01 \times 10^{10}$ vs. $2.64 \times 10^9$ $\Omega \cdot cm$), but the former is denser (4.71 vs. 4.29 g/cm$^3$) with larger grains than the latter (the insets of Fig. 1). So microstructure has a less pronounced effect on humidity sensitivity for $n$-type strontium-hexaferrite semiconductors, while sample $X_M$ has only one-electron donation by chemisorption and $Z_M$ has two-electron donation. Namely, electronic conduction is dominant for sample $Z_M$ in the entire humidity range. But electronic coupled with protonic conduction for sample $X_M$ is dominant, especially at median and high humidity.

Next sample $X_M$ is characterized by humidity properties, using 100 Hz as the optimum frequency. This is because the operation frequency of a humidity sensor is 100 Hz - 1 kHz in practical applications, and devices working at 100 Hz have less interference (vs. 10 Hz) from circuits.

Adsorption is an exothermic and spontaneous process, while desorption needs external energy to relieve the bonding between water molecules and the ceramic surface. Therefore, it leads to the hysteresis between adsorption and desorption processes[25]. For practical applications a sensor must have minimum hysteresis value. The humidity hysteresis characteristic of sample $X_M$ at 100 Hz is shown in Fig. 3. It can be seen that the impedance $Z$ in desorption process is smaller than that in the adsorption. A maximum humidity hysteresis of ~3% RH is observed under 30% RH. The response time for the response and recovery behavior is defined as the adsorption time to reach 90% of the total impedance change when a sensor is switched from 11% to 98% RH, and the recovery time corresponding to the case of desorption from 98% to 11% RH[25]. Fig. 4 shows the response and recovery characteristic curve of sample $X_M$ at 100 Hz. It can be observed that the response time is ~1 s, the equilibrium time is ~150 s, and the recovery time is ~1 s.

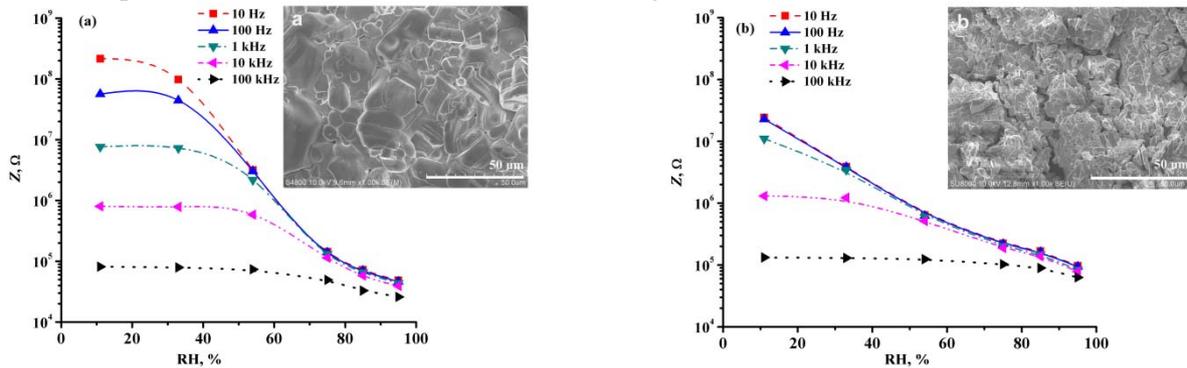

Fig. 1. The dependence of impedance ($Z$) on RH for sample $X_M$ (a) and sample $Z_M$ (b) at 10 Hz - 100 kHz. The insets are the corresponding SEM imagines.

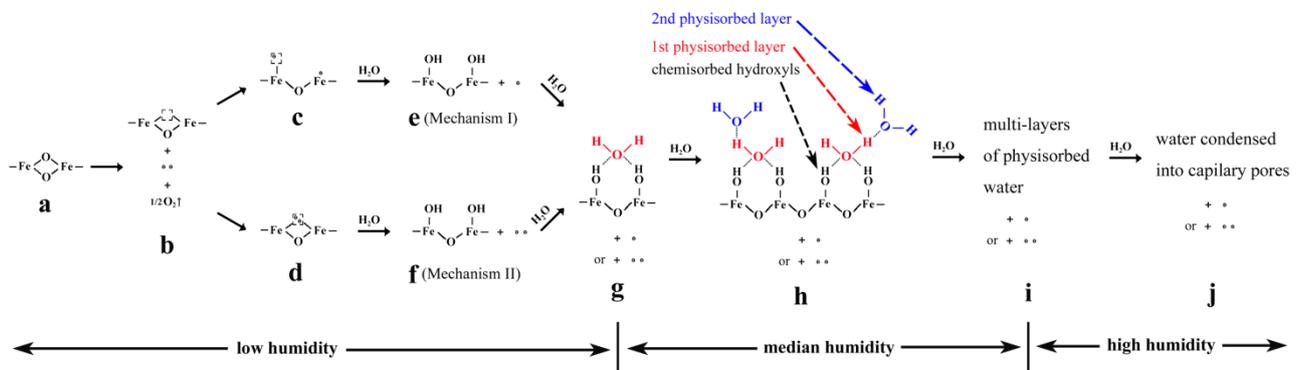

Fig. 2. Contact and reactions between water and semiconductor surface at different humidities. The formation of oxide ion vacancy ($V_{\ddot{O}}$) (a, b); the formation of singly-ionized ($V_{\dot{O}}$) / fully-ionized oxide ion vacancy ($V_O^{\times}$) (c, d); the chemisorption to provide one-electron (where $Fe^{2+}$ generated) (e) or two-electron donation (f); the 1$^{st}$ (g)[19] and 2$^{nd}$ (h)[20] physisorbed layers on the chemisorbed $OH^-$ layer; the multi-layer of physisorbed water (i); and the water condensed into capillary pores (j).

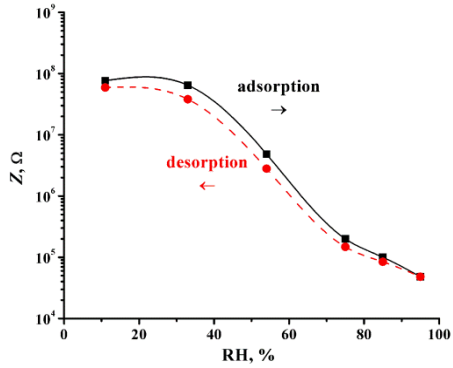 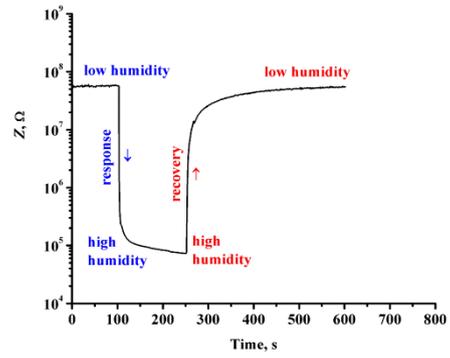

Fig. 3. The humidity hysteresis of sample $X_M$ at 100 Hz. The adsorption process is from low to high RH, and the desorption process is from high to low RH.

Fig. 4. The response and recovery characteristic curve of sample $X_M$ at 100 Hz.

### 3.2. Conduction and humidity sensing mechanisms

Calculated from the synthesized $Sr_2Co_2X$[17] and $Sr_3Co_2Z$[18] hexaferrites by our group, the distance between Fe(Co)-Fe(Co) ions at octahedral sites is shorter than that at octahedral - tetrahedral sites (in S-block for $Sr_2Co_2X$, and in S- / T- blocks for $Sr_3Co_2Z$). And the latter is shorter than that at octahedral - trigonal bipyramidal sites (in R-block for $Sr_2Co_2X$ and $Sr_3Co_2Z$). Therefore hopping between octahedral and trigonal-bipyramidal sites or between octahedral and tetrahedral sites has very small probability compared with that for octahedral-octahedral hopping. On the other hand, the tendency of $Fe^{3+} \rightarrow Fe^{2+}$ in cobalt containing ferrites would be reduced by the reaction $Fe^{2+} + Co^{3+} \rightarrow Fe^{3+} + Co^{2+}$[9, 33]. All of $Co^{2+}$ ions are in S-block for $Sr_2Co_2X$ (there are no T-block in it), but most of $Co^{2+}$ ions are in T-block while a little in S-block for $Sr_3Co_2Z$. Furthermore, the distance between Fe(Co)-Fe(Co) ions at octahedral sites in S-block is shorter than that in T-block for $Sr_3Co_2Z$. So the conduction mechanism in hexaferrites can be explained on the basis of electrons hopping between $Fe^{3+} \leftrightarrow Fe^{2+}$ at octahedral sites, i.e. $Fe^{3+} + e^- \rightarrow Fe^{2+}$, $O^{2-} + 2e^- \rightarrow O$[12]. The difference of conductivity between X- and Z-type hexaferrites is due to that all of $Co^{2+}$ ions are in S-block for $Sr_2Co_2X$ but most in T-block for $Sr_3Co_2Z$.

Electrochemical impedance analysis is used to study humidity sensing mechanism. The equivalent circuit of $R_I(((C_G[R_GW_s])[R_{GB}C_{GB}])[R_WC_W])$ is modeled by Zview software (the inset of Fig. 5a) based on the work of Jansen et al.[34-36], Beekmans et al.[37] and Li et al. (this group)[3]. $R_I$ represents the resistance of interface between the sensing material and Ag electrodes, $R_G/C_G$ represent the resistance and capacitance of interior material, $W_s$ represents the Warburg response related to semi-infinite diffusion of Ag electrodes and electrolytic conduction aroused by water condensed inside pores of the sensing material, $R_{GB}/C_{GB}$ represent the resistance and capacitance of grain boundary, $R_W/C_W$ represent the resistance and capacitance of absorbed water on the sensing surface, i.e., corresponds to the $H_3O^+$ and proton ($H^+$) conductivity, respectively. The complex impedance (Z) is resolved into real ($Z_{re}$) and imaginary ($Z_{im}$) parts, i.e. $Z = Z_{re} + jZ_{im}$. The complex modulus (M) is into $M_{re}$ and $M_{im}$, i.e. $M = j\omega C_0 Z = -\omega C_0 Z_{im} + j\omega C_0 Z_{re} = M_{re} + jM_{im}$. Here the angular frequency $\omega = 2\pi f$ (s$^{-1}$), $C_0$ is the geometrical capacitance (= $\varepsilon_0 S/l$), $\varepsilon_0$ the permittivity of free space (= $8.854 \times 10^{-12}$ F·m$^{-1}$), S the electrode area and l the specimen thickness)[38]. Thus the Nyquist / Cole-Cole plots are constructed.

Fig. 5 shows the measured and simulated complex impedance spectra of sample $X_M$ at 10 Hz - 100 kHz and 11% - 95% RH. Fig. 6 shows the dependance of the corresponding imaginary impedance ($Z_{im}$) and imaginary modulus ($M_{im}$) on frequency, respectively. The relaxation peaks / dispersion were assigned respectively to $R_{GB}/C_{GB}$, $R_G/C_G$, $R_W/C_W$ and $W_s$. Fig. 7 shows the corresponding simulated $R_I$, $R_{GB}$, $R_G$ and $R_W$ variation. It can be seen that the proposed model fits the measured data well. The complex impedance at low RH of 11% - 33% describes a depressed arc / semicircle with a large radius of curvature in the whole frequency region, representing a Randles-like behavior with three parallel resistor-capacitor pairs. Each curve at 54% - 95% RH is composed by a depressed semicircle / arc in the high frequency region and a straight line in the low frequency region, indicating a Warburg behavior aroused by water.

At 11% RH, the $Z_{im}$ vs. $f$ spectra had a dispersion below $5 \times 10^1$ Hz, whereas two $M_{im}$ relaxation peaks were observed at $5 \times 10^1$ Hz ($R_{GB}/C_{GB}$) and $4 \times 10^4$ Hz ($R_W/C_W$), and a high-frequency dispersion above $6 \times 10^4$ Hz ($W_s$). At 33% RH, the $Z_{im}$ vs. $f$ spectra had a peak at $10^2$ Hz ($R_{GB}/C_{GB}$), whereas two $M_{im}$ peaks were at $10^2$ Hz ($R_{GB}/C_{GB}$) and $4 \times 10^4$ Hz ($R_W/C_W$), and a dispersion above $6 \times 10^4$ Hz ($W_s$). At 54% RH, the $Z_{im}$ vs. $f$ spectra had a peak at $2 \times 10^3$ Hz ($R_G/C_G$), whereas two $M_{im}$ peaks at $2 \times 10^3$ Hz ($R_G/C_G$) and $4 \times 10^4$ Hz ($R_W/C_W$). At 75 - 95% RH, the spectroscopic plots of $Z_{im}$ and $M_{im}$ showed a dispersion at high frequencies of above $10^5$ Hz ($W_s$).

The depressed semicircle / arc in Fig. 5 indicates a departure from the Debye-like relaxation behavior, that is, there is a distribution of relaxation times (confirmed by Fig. 6) instead of a single relaxation time[23, 39]. The curvature radius decreases with RH increasing (Fig. 5), corresponding to the decreasing of $R_{GB}$ / $R_G$ / $R_W$ (while $R_I$ keeps very small all the way) (Fig. 7). The $C_{GB}$ / $C_G$ / $C_W$ changed slightly in the entire humidity range (not shown here). In polycrystalline ceramic ferrites, low resistance grains are separated by highly resistive grain boundaries. At median and high humidity, water diffusion from the surface into the structure decreases $R_G$ value. Water absorption on the boundary decreases $R_{GB}$ value. Unchanged capacitances ($C_{GB}$ / $C_G$ / $C_W$) and especially the Warburg response ($W_s$) become dominant parameters to influence the complex impedance. As stated before, sample $X_M$ shows less sensitivity at higher frequency. The impedance decreases as the frequency increases. And the impedance difference at low humidity between two working frequencies is larger than that at high humidity. These humidity sensing effects of sample $X_M$ at low humidity and high frequency are corresponding to the relaxation peaks of imaginary modulus ($M_{im}$) observed at $4 \times 10^4$ Hz and 11 - 54% RH (Fig. 6) due to the absorbed water.

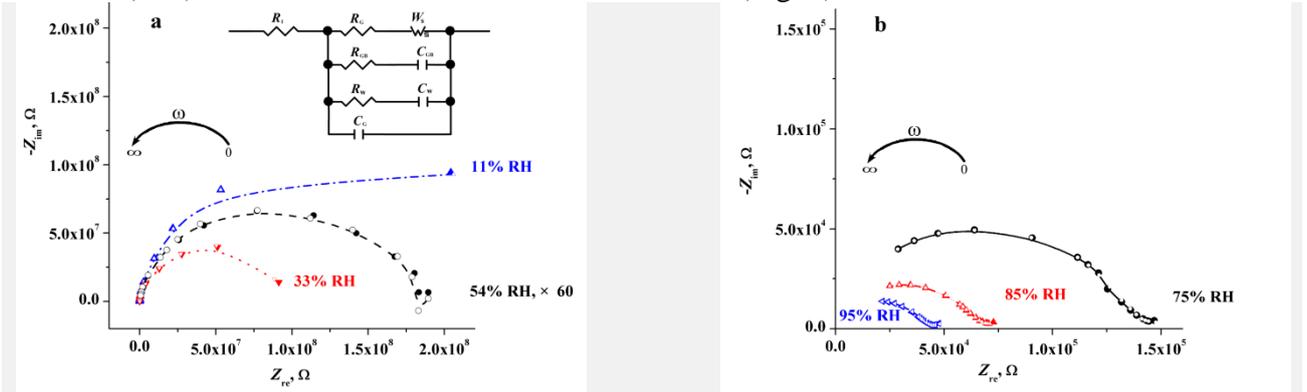

Fig. 5. The measured (solid dots) and simulated (hollow dots and dotted lines) complex impedance spectra of sample $X_M$ at 10 Hz - 100 kHz and 11% - 54% RH (a) / 75% - 95% RH (b). The equivalent circuit is shown in the inset.

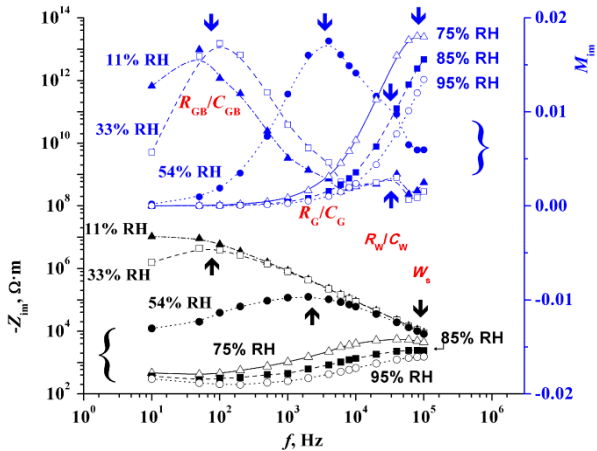

Fig. 6. The measured (dots) and simulated (dotted lines) spectra of sample $X_M$, i.e. $Z_{im}$ vs. $f$ and $M_{im}$ vs. $f$, at 10 Hz - 100 kHz and 11% - 95% RH.

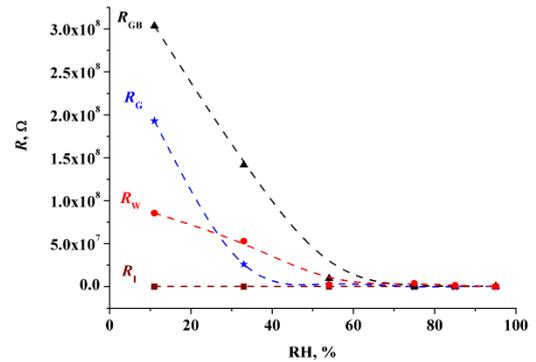

Fig. 7. The variation of $R_I$, $R_{GB}$, $R_G$ and $R_W$ obtained by the simulated data based on sample $X_M$ at 11% - 95% RH.

## 4. Conclusions

The difference of conductivity between X- and Z-type hexaferrites is due to that all of $Co^{2+}$ ions are in S-block for $Sr_2Co_2X$ but most in T-block for $Sr_3Co_2Z$, while the tendency of $Fe^{3+} \rightarrow Fe^{2+}$ would be reduced by the reaction $Fe^{2+} + Co^{3+} \rightarrow Fe^{3+} + Co^{2+}$. Electronic conduction is dominant for Z-type hexaferrite in the entire humidity range. But electronic coupled with protonic conduction for X-type hexaferrite mixture is dominant, especially at median and high humidity where the Warburg response cannot be neglected. The humidity sensing effects of sample $X_M$ at low humidity and high frequency are explained based on the relaxation peaks of imaginary modulus aroused by the absorbed water. The research for the humidity sensitivity of *n*-type hexaferrite semiconductors may give us more enlightenment to magneto-dielectric (MD) / magneto-electric (ME) effects.

**Acknowledgements**

Authors gratefully thank Ms. D.-M. Xu in Elite, China. This work was supported by the Research Foundation of SKLWUT, China (Grant Nos. 2014-KF-6, 2015-KF-4 & 2016-KF-4).